\documentclass[ApJL,twocolumn]{aastex62}
\usepackage{amssymb, amsmath, graphicx, dcolumn, color,units, xspace, mathtools, physics, tensor, bm, lipsum, revsymb}
\usepackage[normalem]{ulem}
\usepackage{hyperref}
\usepackage{booktabs,dcolumn}

\newcommand{\ozstar}{{\sc OzStar}\xspace}

\newcommand{\bilby}{{\sc Bilby}\xspace}
\newcommand{\gwcloud}{{\sc GWCloud}\xspace}

\newcommand{\jupyter}{{\sc Jupyter}\xspace}
\newcommand{\python}{{\sc Python}\xspace}

\newcommand{\ini}{{\tt .ini}\xspace}
\newcommand{\prior}{{\tt .prior}\xspace}
\newcommand{\psd}{{\tt PSD}\xspace}
\newcommand{\calib}{{\tt .calib}\xspace}

\newcommand{\outputdir}{{\tt output/}\xspace}

\newcommand{\SPA}{School of Physics and Astronomy, Monash University, Clayton VIC 3800, Australia}
\newcommand{\OzGravMonash}{OzGrav: The ARC Centre of Excellence for Gravitational Wave Discovery, Clayton VIC 3800, Australia}

\shorttitle{GWCloud}
\shortauthors{Baker \textit{et al}.}

\begin{document}

\title{\gwcloud: a searchable repository for the creation and curation of gravitational-wave inference results}

\author{A. Makai Baker}
\author{Paul D. Lasky}
\author{Eric Thrane}
\affiliation{\SPA}
\affiliation{\OzGravMonash}
%\correspondingauthor{Gwynne C. Loud}
\email{paul.lasky@monash.edu}
\email{eric.thrane@monash.edu}

\author{Gregory Ashton}
\affiliation{Department of Physics, Royal Holloway, University of London, TW20 0EX, United Kingdom}

\author{Jesmigel Cantos}
\author{Lewis Lakerink}
\author{Asher Leslie}
\author{Gregory B. Poole}
\author{Thomas Reichardt}
\affiliation{Astronomy Data and Computing Services (ADACS); the Centre for Astrophysics \& Supercomputing, Swinburne University of Technology, P.O. Box 218, Hawthorn, VIC 3122, Australia}

\date{\today}

\begin{abstract}
% Eric: this includes 90 identified by LIGO + 10 claimed by IAS in arxiv/2201.02252.
There are at present ${\cal O}(100)$ gravitational-wave candidates from compact binary mergers reported in the astronomical literature.
% Eric: reference for the next sentence: https://arxiv.org/pdf/1906.04197.pdf
As detector sensitivities are improved, the catalog will swell in size: first to ${\cal O}(1000)$ events in the A+ era and then to ${\cal O}(10^6)$ events in the era of third-generation observatories like Cosmic Explorer and the Einstein Telescope.
Each event is analyzed using Bayesian inference to determine properties of the source including component masses, spins, tidal parameters, and the distance to the source.
These inference products are the fodder for some of the most exciting gravitational-wave science, enabling us to measure the expansion of the Universe with standard sirens, to characterise the neutron star equation of state, and to unveil how and where gravitational-wave sources are assembled.
In order to maximize the science from the coming deluge of detections, we introduce \gwcloud, a searchable repository for the creation and curation of gravitational-wave inference products.
It is designed with five pillars in mind: uniformity of results, reproducibility of results, stability of results, access to the astronomical community, and efficient use of computing resources.
We describe how to use \gwcloud with examples, which readers can replicate using the companion code to this paper.
We describe our long-term vision for \gwcloud.
\end{abstract}

\section{Introduction}
The recent release of the third LIGO--Virgo--KAGRA (LVK) Gravitational-Wave Transient Catalog (GWTC-3) increased the total number of gravitational-wave events detected by LVK to 90 \citep{gwtc-3}.
\cite{Olsen2022} detect an additional ten events with the so-called IAS pipeline while \cite{Nitz2021} detect seven events not included in GWTC-3.
The GWTC-3 catalog consists mostly of binary black hole mergers with two binary neutron star mergers and $\gtrsim 2$ two neutron star + black hole mergers.\footnote{The event GW190814 \citep{GW190814} could be a binary black hole or a neutron star + black hole binary.}
Each event is characterized by $\approx 15$-$17$ astrophysical parameters \citep{lalinference}.
There are seven extrinsic parameters describing the location and orientation of the event with respect to the observatory and eight or more intrinsic parameters including the component masses and spin vectors.
Tidal parameters are often included for systems with a neutron star candidate \citep{Lackey2015,GW170817_EOS,Chatziioannou2020} and some binary black hole analyses now include parameters characterising orbital eccentricity \citep{eccentricity,gwtc1_eccentricity,GW190521_formation,gayathri22,Lennon2020,gwtc2_eccentricity}.

Each event is analyzed with a Bayesian inference pipeline \citep{lalinference,bilby,bilby_gwtc1,Biwer2019,rift} in order to determine its astrophysical parameters.
The output of these pipelines typically includes posterior samples---discrete representations of the posterior distribution, which can be used to calculate credible intervals for different combinations of astrophysical parameters (corner plots).
However, their usefulness does not end there, as they serve as a reduced data product for some of the most exciting gravitational-wave science.
Posterior samples are used for population studies \citep{intro,Vitale2021} to probe how compact binaries are distributed in mass and spin, providing insights into stellar evolution and binary formation (see, e.g., \cite{gwtc-3_pop,gwtc-3_cosmo,o3a_pop,o2_pop}).
They are  used in standard-siren analyses \citep{Schutz1986,Holz2005} to measure cosmological expansion (see, e.g., \cite{gwtc-3_cosmo}) and to determine the neutron star equation of state \citep{Baiotti2019,Lackey2015,Landry2019,stacking2,Wysocki2020,lvk_eos}.

While the output of inference pipelines is crucially important to gravitational-wave astronomy, there are several aspects of these inference products that make life complicated for gravitational-wave astronomers.
First, it is often necessary to carry out many different inference calculations, which analyse the same event with subtle differences.
In particular, there are several popular waveform approximants used to model gravitational waveforms, each with different capabilities and different systematic errors.
It is therefore common to analyse each event with multiple waveforms.
Likewise, it is sometimes necessary to analyse the same event with different prior assumptions (for example, assuming one or both compact objects are not spinning \citep{BuildingBetterModels}), which can also lead to multiple runs.
There may also be runs carried out with different samplers or different sampler settings, which occasionally yield qualitatively different results (see, e.g., \citep{Chia2021,deep}).
Finally, the data itself can have different versions due to variations in calibration \citep{Sun2020}, cleaning \citep{gwosc2021}, and/or glitch subtraction schemes \citep{Chatziioannou2020b}.

The large number of inference results associated with each event creates a book-keeping problem.
This problem is compounded by a second issue: the high computational cost of gravitational-wave inference.
Even a ``fast'' run on an ordinary binary black hole event with a cheap approximant can take $\approx\unit[10]{hrs}$.
More ambitious analyses on longer signals and/or with cutting-edge approximants can take weeks.
Given the substantial cost in time and CO$_2$ emission associated with the generation of astrophysical inference results, it is becoming increasingly necessary to carefully curate gravitational-wave inference results.

Finally, the lack of a centralized repository for inference results makes the current workflow of gravitational-wave astronomers inefficient and susceptible to error.
A researcher looking for the output of a particular inference run, (e.g., using the \textsc{IMRPhenomXPHM}approximant \citep{Pratten2021} to analyze GW151226 \citep{GW151226} with special sampler settings) may need to email collaborators to find the results.
The results in question may have been subsequently moved or even deleted.
And the researcher cannot be certain that the files she tracks down are precisely what she is looking for.

The situation is already challenging with 90 events.
The difficulties will, of course, increase as the gravitational-wave catalog swells \citep{Baibhav2019} to ${\cal O}(1000)$ events in the A+ era and ${\cal O}(10^6)$ events in the era of third-generation observatories such as Cosmic Explorer \citep{Reitze2019,Evans2021} and the Einstein Telescope \citep{Maggiore2020}.
In order to address these challenges, we introduce \gwcloud, a searchable repository for the creation and curation of gravitational-wave inference results.
There are five pillars underpinning its design philosophy:\footnote{These design pillars are aligned with the Australian Research Data Commons guidelines for ``FAIR'' data, which is findable, accessible, interoperable, and reusable. For more information, see \url{https://ardc.edu.au/resources/aboutdata/fair-data/}.}
\begin{enumerate}
    \item \textbf{Uniformity of results.}
    Inference results are downloaded and uploaded in a uniform format.
    Uniformity facilitates validation:  new results must pass checks to ensure that the inference output is complete and uncorrupted with the necessary metadata to repeat the analysis.
    %%%
    \item \textbf{Reproducibility of results.}
    By curating the metadata and code version of each result, we insist that every entry in \gwcloud can be reproduced.
    %%%
    \item \textbf{Stability of results over time.}
    Each result is assigned a permanent location.
    Users can locate previous results using a search engine.
    Before launching a new inference job, users can search to see if the analysis they want has already been performed.
    Avoiding duplicate analyses reduces the carbon footprint of gravitational-wave astronomy.
    %%%
    \item \textbf{Access to results.}
    While a large fraction of gravitational-wave astronomy effort takes place within the LVK collaboration, significant advances are now made by external groups.
    \gwcloud provides multiple levels of access so that results can be shared both within the LVK collaboration and to the larger astronomical community, facilitating the exchange of ideas among a broad community.
    %%%
    \item \textbf{Efficient use of computing resources.}
    \gwcloud enables users to submit inference jobs on multiple computing clusters through a single portal.
    Each cluster can use different batch queuing protocols (e.g., \textsc{slurm} versus \textsc{condor}) and allow for different user groups (e.g., LVK users versus the general public).
    In this way, \gwcloud helps match users with computing resources.
\end{enumerate}

\gwcloud is not the only tool that has been created to tackle these challenges.
The Gravitational-Wave Open Science Centre (GWOSC) provides access to most of the publicly available posterior samples used in LVK papers.
The samples can be queried, discovered, and downloaded through the GWOSC Event Portal, at \url{https://gwosc.org/eventapi}.
These samples are also available through \url{zenodo.org}.
Recently, \cite{Asimov} introduced \textsc{Asimov}, a framework for coordinating parameter estimation workflows.
It includes a number of useful features, including a review sign-off system so that key results are vetted by humans.
Meanwhile, the program \textsc{PESummary} \citep{pesummary} has helped facilitate the dissemination of uniform results (while simultaneously providing a tool for the visualisation of inference results).
It includes functionality to access result files\footnote{\url{https://lscsoft.docs.ligo.org/pesummary/stable_docs/gw/fetch.html}} (both public and private) and functionality to reproduce results\footnote{\url{https://lscsoft.docs.ligo.org/pesummary/stable_docs/gw/cli/summaryrecreate.html}}.
\textsc{PESummary}, \textsc{Asimov}, and \gwcloud provide complementary services, although the way in which they will interact in the future is not yet clear.

The remainder of this paper is organized as follows.
In Section~\ref{basics}, we cover the basics of \gwcloud: how to submit new inference jobs and how to upload the results of an inference analysis.
In Section~\ref{GW150914}, we provide the first of three case studies: we use \gwcloud to reanalyze the iconic event GW150914 \citep{GW150914}, but with the assumption that both black holes have negligible spin.
In Section~\ref{correlations}, we present the second case study: using \gwcloud to investigate correlations between mass and spin parameters using events in the second gravitational-wave transient catalog (GWTC-2) \citep{gwtc-2}.
In Section~\ref{GW190521}, we describe the third case study: using \gwcloud to download posterior samples for the remarkably high-mass event GW190521 \citep{GW190521} obtained using an eccentric waveform approximant.
We conclude in Section~\ref{conclusions} with a discussion of future development plans.
Technical details are provided in the appendix (Section~\ref{appendix}).
The case studies presented in this paper are supported by \jupyter notebooks available as part of the online supplement here: \url{https://git.ligo.org/gwcloud/paper/}.

\section{Basics}\label{basics}
\subsection{What is \gwcloud?}
In this Section, we provide a high-level overview of \gwcloud and instructions for the most basic tasks users are likely to perform.
\gwcloud consists of two components: a portal to launch inference jobs and a database to store the results of inference jobs.
Both components can be accessed using a web-based graphical user interface (UI) at \url{https://gwcloud.org.au/}.
Users who prefer to access \gwcloud entirely with command-line programming may instead use the application programming interface (API).
We anticipate that the UI's job submission feature will be most useful for new and casual users.
However, the UI's search feature should be useful to any user searching for old inference results.
The API is likely to be most useful to experts who sometimes need to submit large batches of jobs.
It allows for more complicated job submissions with features that are not supported using the UI (for example, custom priors).

The portal launches inference jobs using \bilby \citep{bilby,bilby_gwtc1}.
Jobs launched through the UI are at present run on the \ozstar cluster based at Swinburne University.
Jobs submitted by authenticated LVK users through the API can also be run on computers that form part of the LIGO Data Grid.
It is also possible to upload jobs to the \gwcloud database that were not run through the \gwcloud portal so long as they are in the standard \bilby format (see Section \ref{upload}).\footnote{This feature can be used to upload results from other inference code such as \textsc{LALInference} \citep{lalinference} or \textsc{RIFT} \citep{rift}.}
This is useful for storing jobs that were run before the creation of \gwcloud or jobs that require special resources to run, for example, computationally expensive Parallel \bilby~\citep{pBilby} analyses that require a high-performance computing cluster.

\begin{figure*}
    \centering
    \fbox{
    \includegraphics[width=\textwidth]{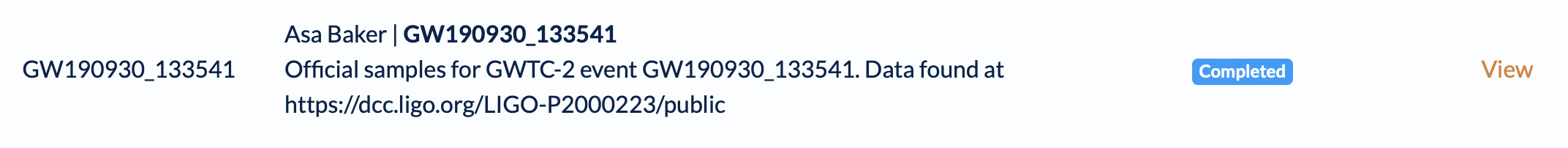}
    }
    \caption{
    Example entry in \gwcloud.
    }
    \label{fig:example_entry}
\end{figure*}

Users visiting the \gwcloud landing page are met with a prompt requiring sign-in.
Members of the LVK collaboration can sign in using their \texttt{albert.einstein} credentials, which also provides access to the LIGO Data Grid, while other users can create a \gwcloud account.
After logging in, the user is taken to the ``public jobs'' page, which lists the most recent \gwcloud runs; see Fig.~\ref{fig:example_entry} for an example entry from this recent-job list.
The search field allows users to find jobs based on their \texttt{description}, the \texttt{user} who submitted the job, the \texttt{job\_name}, and the \texttt{event ID}.
\texttt{Labels} are available to distinguish some jobs as special.
For example, the \texttt{preferred} label indicates that a job is used for an official LVK result.\footnote{Other currently available labels include \texttt{Bad run}, \texttt{Production run}, \texttt{Review requested}, and \texttt{Reviewed}.}
Previous jobs can be viewed and downloaded by clicking on the appropriate \texttt{view} link. 
Users may create a new job by clicking on \texttt{start a new job} and following the instructions.

In the next subsections we describe how to submit (or upload) a job using the API.
Additional information is provided on the \gwcloud web page by clicking on \texttt{Python API.}
In order to implement the examples below, readers must install the \gwcloud API:
\begin{verbatim}
    pip install gwcloud-python
\end{verbatim}

\subsection{Submitting a new job with the \gwcloud API}\label{submit}
Here we describe a \python script for submitting a new \gwcloud job using the API; see
\sloppy
\texttt{JobSubmission.ipynb} for the corresponding \jupyter notebook.
The corresponding job can be viewed on the \gwcloud UI by searching for the name: \texttt{GW150914Example}.

The first step is for the user to authenticate by initialising a token identifier generated by \gwcloud. 
At the beginning of any \gwcloud script, include the following lines to import the \gwcloud API and set up your token:
\begin{verbatim}
    from gwcloud_python import GWCloud
    gwc = GWCloud(token=`YourTokenHere') 
\end{verbatim}

The next step is to create a \bilby \ini file, which is required to submit a job with \gwcloud because it is required to run \bilby.
The \ini file for this tutorial is \texttt{GW150914\_example.ini}.
The \ini file tells \bilby which data to analyze and how to analyze it.\footnote{
Please see \url{https://lscsoft.docs.ligo.org/bilby/} for additional \bilby documentation.}
The \ini file contains local paths to noise power spectral density (\psd) file(s), spline calibration (\calib) file(s), and the \prior file.
All of these files are uploaded to \gwcloud for reproducibility.
With the \ini file ready, we submit the job to the LIGO Data Grid's Caltech cluster like so:
%%%%%%%%
\begin{verbatim}
	new_job = gwc.start_bilby_job_from_file(
    job_name = ``GW150914Example'',
    job_description = ``Testing GWCloud'',
    private=False,
    ini_file=`GW150914_example.ini',
    cluster = Cluster.CIT)
\end{verbatim}
Once this command is executed, a new job with \texttt{job\_name = GW150914Example} becomes visible on the \gwcloud UI.\footnote{The job name, combined with the \gwcloud user name of the person who submitted the job uniquely define each event. Thus, two different users can have a job named \texttt{GW150914Example}, but one user can not give this name to two different jobs.}
Since we set \texttt{private=False}, the job can be viewed by anyone using \gwcloud.\footnote{Jobs are marked as \texttt{LVK} or not. They are also marked as \texttt{private} or not. A job with \texttt{LVK=true} and \texttt{private=false} may be viewed by all members of the LVK Collaboration.}
The last line of code tells \gwcloud to run this job on the Caltech computing cluster.

The progress of the new job can be monitored with the \gwcloud UI.
When the job is complete, the API can be used to retrieve the posterior samples from \gwcloud with the following command:
\begin{verbatim}
    job.save_result_json_files(`/path/')
\end{verbatim}
Which saves the result files containing the posteriors to the specified path.

\subsection{Uploading the results of an existing \bilby run}\label{upload}
Here we describe a \python script to upload existing results to \gwcloud job using the API; see \texttt{JobUpload.ipynb} for the corresponding \jupyter notebook.
The corresponding job can be viewed on \gwcloud by searching for \texttt{job\_name = GW190412} by \texttt{user = Asa Baker}.
As our starting point, we need a \bilby{} \outputdir directory with the requisite subdirectory structure.
We modify the \texttt{label} field in the \texttt{*\_config\_complete.ini} file to set the \gwcloud job name, e.g.,
\begin{verbatim}
    label = `GW150914_Upload_Example'
\end{verbatim}
Next, create a tar-zipped file of the \bilby{} \outputdir directory, which can be accomplished by running this command:
\begin{verbatim}
    tar -cvf archive.tar.gz .
\end{verbatim}
Finally, the job is submitted by uploading the tar-zipped file to \gwcloud: 
\begin{verbatim}
    gwc.upload_job_archive(‘Example upload
        with GW159014.’, 
        ‘/path/archive.tar.gz’)
\end{verbatim}
\gwcloud checks the submission to make sure all the requisite results and supporting files are included.

\section{Case Study I: submit a job to analyze GW150914 with a zero-spin prior}\label{GW150914}
The \gwcloud graphical UI allows users to submit inference jobs with various default prior settings.
While these settings are probably adequate for new users, expert users will need the API in order perform runs with custom priors.
Here we provide an example of how the API can be used to carry out an inference calculation with a non-standard prior; see \texttt{CaseStudy1.ipynb} for the corresponding \jupyter notebook.
The corresponding jobs can be viewed on \gwcloud by searching for the names: \texttt{GW150914Example} and \texttt{GW150914NoSpin} by \texttt{Asa Baker}.
Specifically, we reanalyze the iconic first binary black hole event GW150914 \citep{GW150914}, but assuming that both black holes have negligible dimensionless spins $\chi_1 = \chi_2 = 0$ (here the $1$ subscript refers to the more massive ``primary'' black hole while the $2$ subscript refers to the less massive ``secondary'' black hole).
This example is motivated by work by \cite{Fuller2019,Miller2020,Roulet2020,BuildingBetterModels,Hoy2022}, which suggest a sub-population of LVK detections are likely characterized by negligible black-hole spin.

\begin{figure*}
    \centering
    \includegraphics[width=0.75\textwidth]{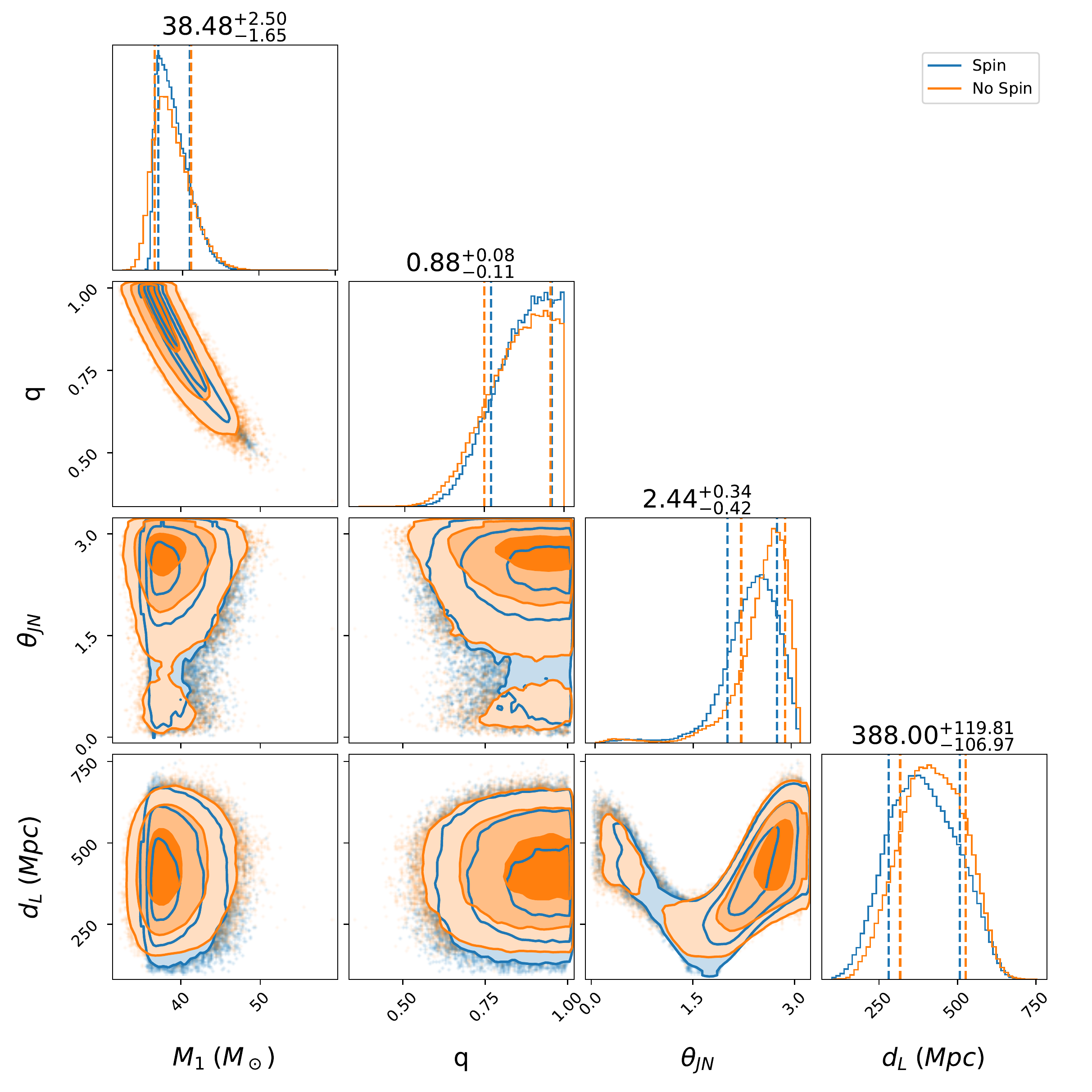}
    \caption{
    A corner plot showing the marginalised posterior distribution of the first binary black hole event GW150914.
    The masses are given provided in the lab frame.
    The default results (calculated with a $U(0,1)$ prior for the dimensionless spins $\chi_1, \chi_2$) is shown in blue while the orange shows the results assuming $\chi_1=\chi_2=0$.
    The different shades indicate one-, two-, and three-sigma credible intervals.
    }
    \label{fig:GW150914}
\end{figure*}

We prepare two \ini files: \texttt{GW150914.ini} reproduces standard \bilby settings for a short-duration (high-mass) binary black hole signal.
The prior for the dimensionless spins $\chi_1, \chi_2$ is uniform on the interval of zero to one.
Meanwhile, in \texttt{GW150914\_nospin.ini}, we set $\chi_1=\chi_2=0$.
We submit both jobs and download the results using the syntax described in Section~\ref{submit}.
In Fig.~\ref{fig:GW150914} we provide a corner plot comparing the credible intervals for various parameters of GW150914 assuming a uniform prior for the dimensionless spins (blue) and a no-spin prior (orange).
The different shading indicates one-, two-, and three-sigma credible intervals.
The different choice of prior yields subtle but interesting shifts in the posterior distribution.
Comparing the marginal likelihoods for each run, we find that the zero-spin hypothesis is preferred with a Bayes factor of $\text{BF}=3.7$, consistent with the conclusions from \cite{Miller2020,Roulet2020,BuildingBetterModels,Hoy2022} that some gravitational-wave events are best described as having negligible spin.

\section{Case Study II: Download results from GWTC-2 for Correlation study}\label{correlations}
\begin{figure*}
    \centering
    \includegraphics[width=0.7\textwidth]{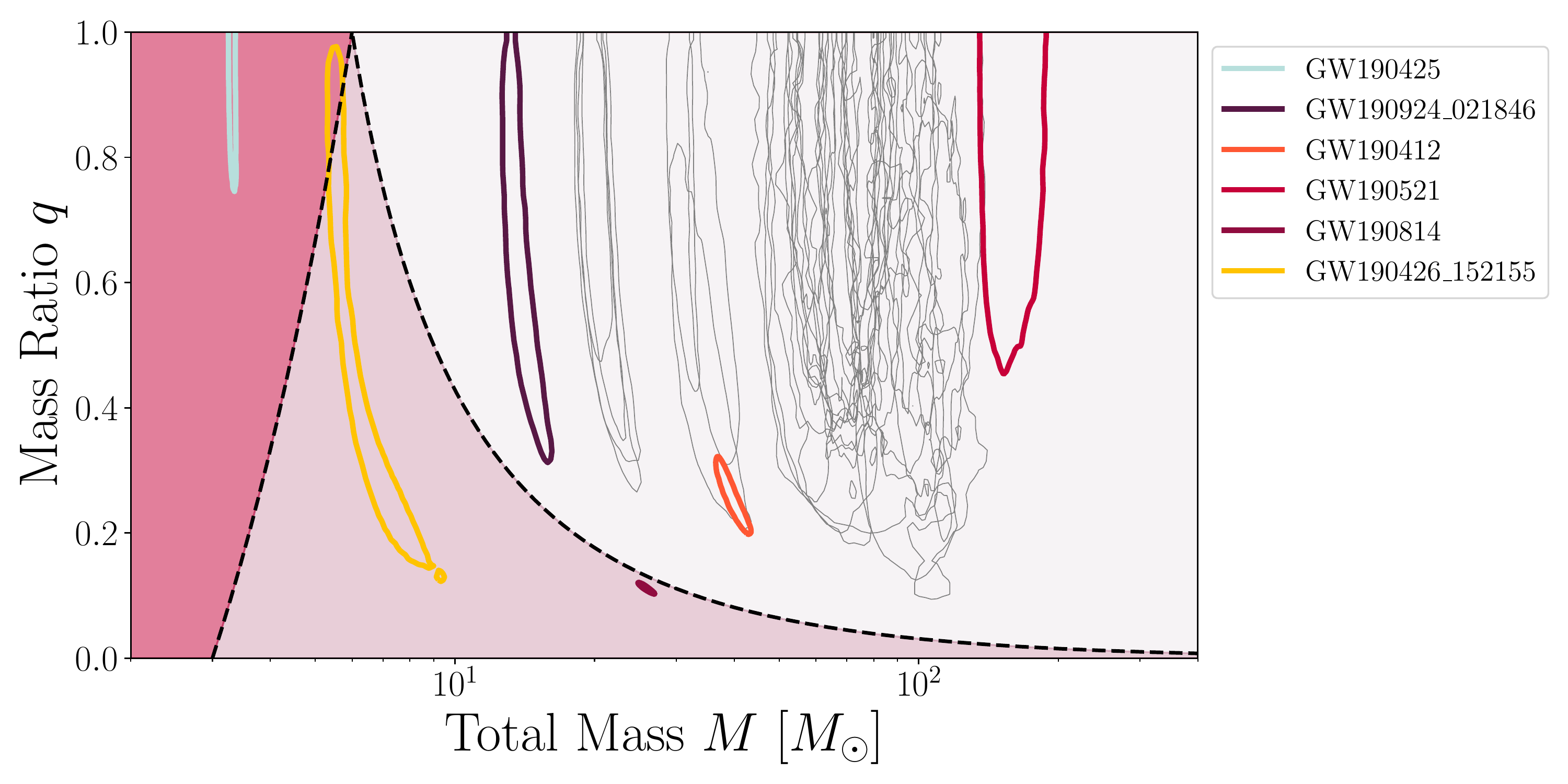}
    \caption{
    Compact binary coalescence events from GWTC-2 in the plane of total mass $M$ and mass ratio $q$. 
    Each contour represents the 90\% credible region.
    Select events are highlighted.
    The dashed lines mark the border beyond which one or more component has a mass $<\unit[3]{M_\odot}$; objects below this threshold are neutron-star candidates.
    The events in the grey region are confidently binary black hole events while events in the mauve region may contain a neutron star.
    The purple region is forbidden by the requirement that $m_1 > m_2$.
    }
    \label{fig:GWCloud_GWTC_masses}
\end{figure*}

\begin{figure*}
    \centering
    \includegraphics[width=0.7\textwidth]{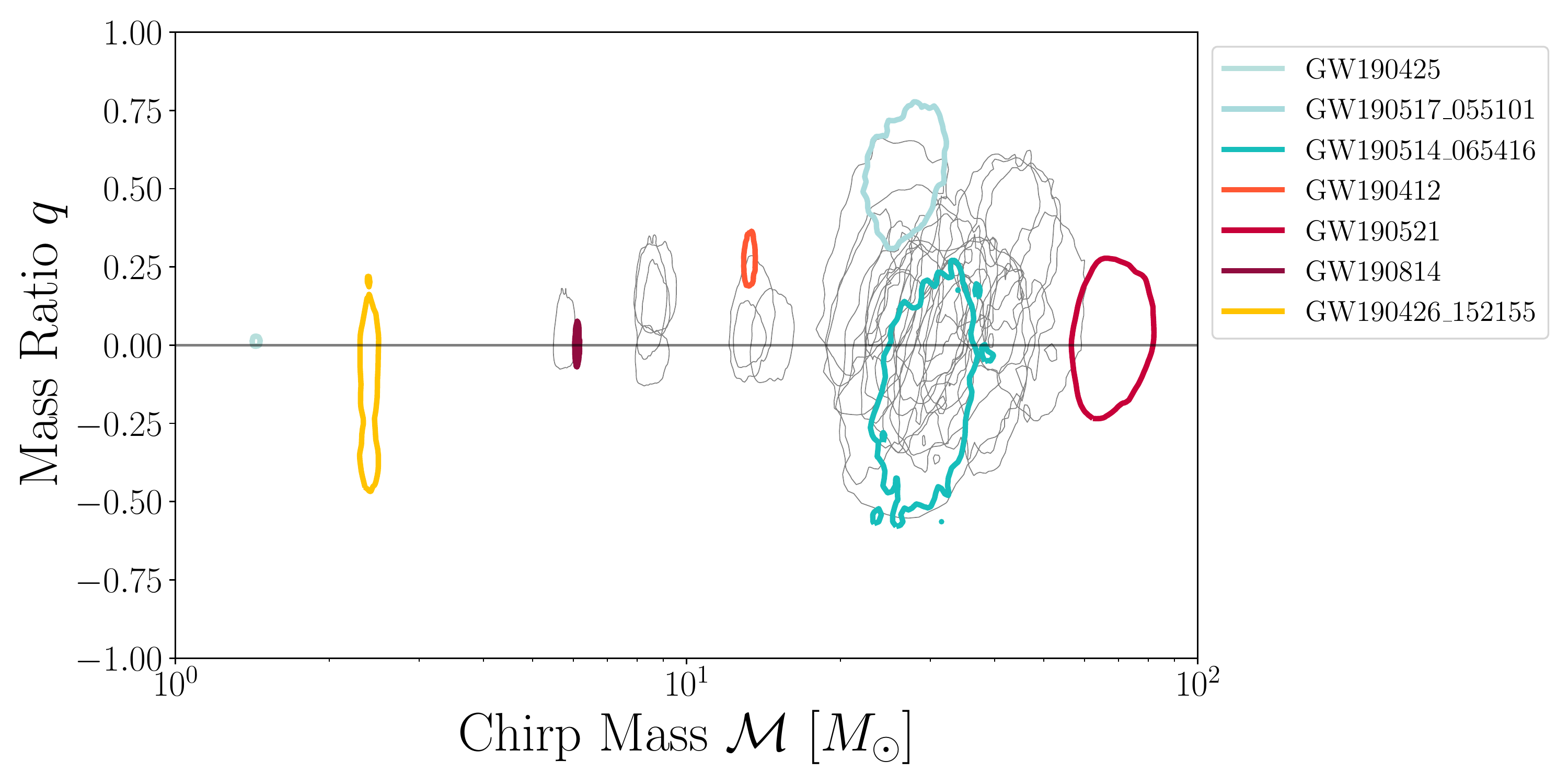}
    \caption{
        Compact binary coalescence events of the LVK GWTC-2 catalog in the plane of chirp mass $\mathcal{M}$ and effective inspiral spin $\chi_\text{eff}$.
        Each contour represents the 90\% credible region for a different event.
        Select events are highlighted.
    }
    \label{fig:GWCloud_GWTC_chieff}
\end{figure*}

In this case study, we provide an example of how the API can be used to download previous inference results to look for trends in the population of merging binary black holes; see \texttt{CaseStudy2.ipynb} for the corresponding \jupyter notebook.
The corresponding jobs can be viewed on \gwcloud by searching for the keyword: \texttt{GWTC-2}.
This example is motivated by work by \cite{Callister2021}, suggesting that black-hole spin is correlated with mass ratio.
We download the ``preferred samples'' (used for official LVK analyses) for 47 binary black hole events in GWTC-2 \citep{gwtc-2,o3a_pop}.
To retrieve these GWTC-2 jobs, we run the following command:
\begin{verbatim}
    jobs = gwc.get_public_job_list(
            search="GWTC-2",
            time_range=TimeRange.ANY)
\end{verbatim}
%To retrieve the jobs labelled as \texttt{preferred}, we run the following command:
%\begin{verbatim}
%    gwc.get_preferred_job_list()
%\end{verbatim}
In Fig.~\ref{fig:GWCloud_GWTC_masses}, we plot the 90\% credible intervals in the plane of total mass $M$ and mass ratio $q$ for events in GWTC-2.
In Fig.~\ref{fig:GWCloud_GWTC_chieff}, meanwhile, we plot credible intervals in the plane of chirp mass $\mathcal{M}$ and the effective inspiral spin $\chi_\text{eff}$.
These two plots can be compared to Figs.~6-7 in \cite{gwtc-2}.
By examining the distributions of events in two-dimensional planes, it is sometimes possible to see previously unknown correlations.
In this case, there is not an obvious correlation present in either plot.

%The colour scheme of these plots may be too monotone etc.

\section{Case Study III: Download results for eccentric analysis of GW190521}\label{GW190521}
Binary black holes formed from stellar binaries are expected to merge with quasi-circular orbits.
However, a non-zero eccentricity may indicate that the binary was assembled from previously unbound black holes, a process called ``dynamical formation.''
We consider GW190521~\citep{GW190521}, one of the most massive binary black hole events to date, which shows signs of non-zero spin precession and/or eccentricity~\citep{GW190521_formation, gayathri22}.
\cite{GW190521_formation} analyzed this event using quasi-circular and eccentric waveforms.
The results of this analysis have been uploaded to \gwcloud and can be viewed on by searching for: \texttt{job\_name = GW190521, user = Asa Baker} (for results obtained with the circular waveform obtained with \textsc{NRSur7dq4} \cite{Varma2019}) and/or \texttt{job\_name = GW190521\_eccentric, user = Asa Baker} (for the results obtained with \textsc{SEOBNRE} \citep{Cao2017,Liu2020}). 

\begin{figure*}
    \centering
    \includegraphics[width = 0.6\textwidth]{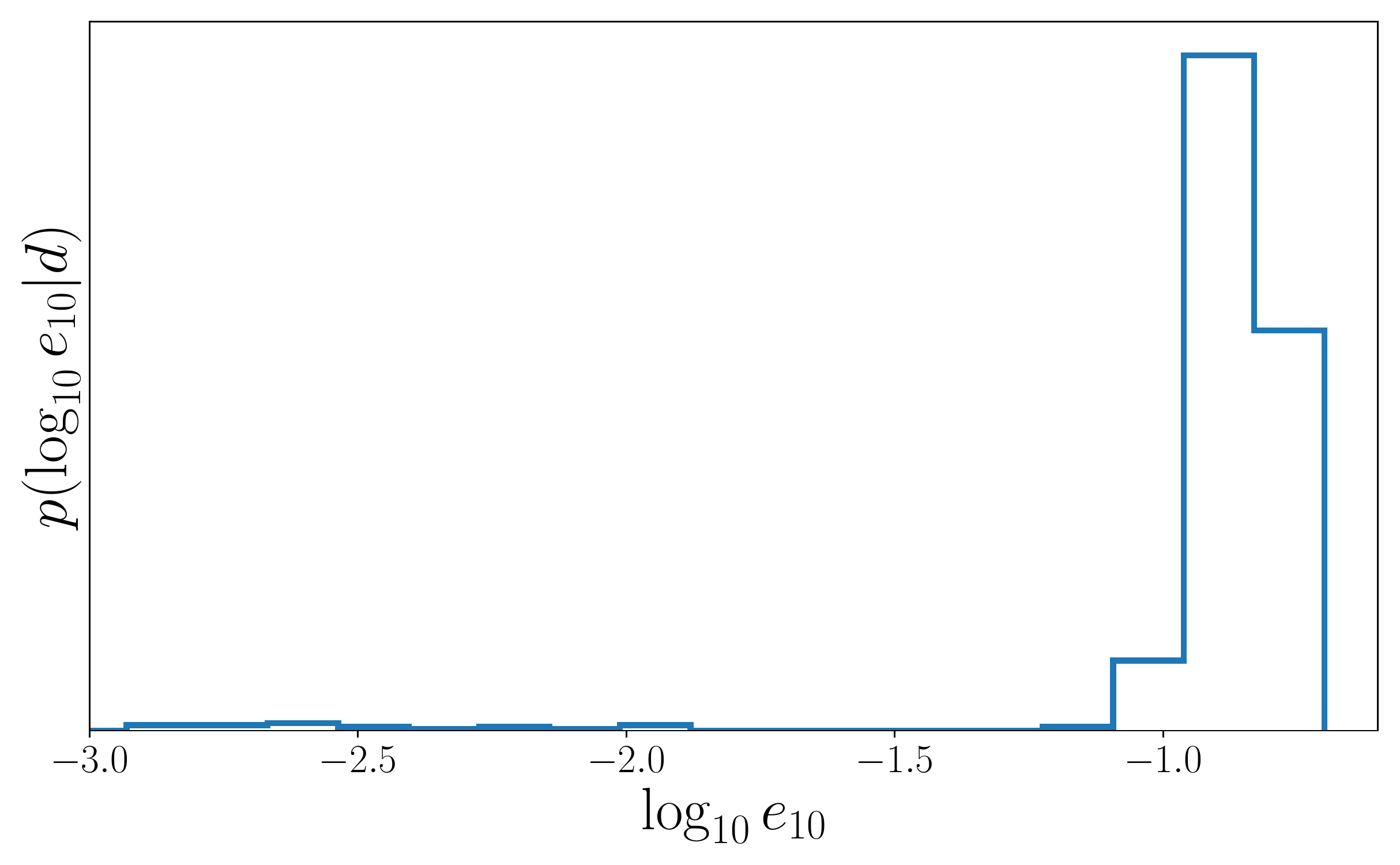}
    \caption{The posterior distribution for the eccentricity of GW190521 at a reference frequency of $\unit[10]{Hz}$ obtained with the \textsc{SEOBNRE} waveform \citep{Cao2017,Liu2020}) by \cite{GW190521_formation}.
    }
    \label{fig:GW190521}
\end{figure*}

To retrieve these jobs from \gwcloud, search for \texttt{GW190521} in the public jobs by executing the following command
\begin{verbatim}
    jobs = gwc.get_public_job_list(
            search="GW190521", 
            time_range=TimeRange.ANY)
\end{verbatim}
and download the jobs by \texttt{Asa Baker}.
In Fig.~\ref{fig:GW190521}, we plot the posterior distribution for the eccentricity of GW190521 at a reference frequency of $\unit[10]{Hz}$ (compare with Fig. 1 of~\citet{GW190521_formation}).
See \texttt{CaseStudy3.ipynb} for the corresponding \jupyter notebook.

\section{Future development}\label{conclusions}
We close by considering the future of \gwcloud, describing new functionality we hope to add in both the short term and long term.
As we plan for the future, we invite input from the astronomical community; please visit our git issue tracker to leave a suggestion or to propose a new feature.\footnote{\url{https://gitlab.com/CAS-eResearch/GWDC/projects/gwcloud/issues}}

\textbf{Short-term goals.}
\begin{enumerate}
    \item \textit{Making LVK jobs public.}
    When LVK data is published, jobs that are previously marked as \texttt{LVK} can be changed to \texttt{public}.
    %%%
    \item \textit{\gwcloud teams.}
    Share jobs among a small team.
    Team members can add comments to different jobs, e.g., ``this result does not look fully converged.''
    Teams can combine jobs to create catalogs.
    %%%
    \item \textit{Archiving complementary information.}
    Gravitational-wave inference results do not exist in vacuum.
    In order to generate and interpret them, we rely on a number of other data products including estimates of the noise power spectral density (e.g., \cite{BayesLine}), injection studies used to quantify selection effects \citep{Talbot2022,Gerosa2020}, and probabilities that a given event is astrophysical $p_\text{astro}$ \citep{Kapadia2020}.
    We hope to extend \gwcloud to include these and other data products.
    %%%
    \item \textit{Visualization.}
    Static and dynamic visualization of inference products is useful to understand covariances.
    Such functionality is currently offered within the \texttt{pe\_summary} toolkit~\cite{pesummary}; a short-term goal is full integration of these visualisation toolkits into the \gwcloud workflow.
\end{enumerate}

\textbf{Long-term goals.}
\begin{enumerate}
     \item \textit{Identify similar jobs.}
    Warn users if they are about to launch a job that is similar to one already in the database.
    Users may choose to use existing results rather than waiting for new ones (and potentially generating more CO$_2$ emissions).
    In some cases, importance sampling can be used to re-weight posterior samples to convert the results from a ``proposal'' distribution to a ``target'' distribution \citep{hom}.
    %%%
    \item \textit{Estimate job run time.}
    Use machine learning to provide estimated time to completion for new jobs.
    Warn users if they launch a job that is likely to take more than a week to complete.
    %%%
    \item \textit{Connecting to other clusters.}
    Currently, \gwcloud provides users access to the computing clusters of the LIGO Data Grid and the OzStar clusters at the Swinburne University of Technology.
    However, \gwcloud could be connected to other computing resources such as the Open Science Grid \citep{Pordes2007}.
    %%%
    \item \textit{Automated inference.}
    The project could be extended to launch automated inference jobs for promising triggers by e.g., integrating with \textsc{Asimov}~\cite{Asimov}.
    When extra computational resources are available, carry out inference on all data segments.
    The results can be used to carry out a statistically optimal search for the astrophysical background \citep{tbs} and to construct fully Bayesian detection statistics \citep{VeitchVecchio,bcr,Pratten2021b}.
    %%%
    \item \textit{Beyond posterior samples.}
    The majority of gravitational-wave inference relies on posterior samples.
    However, in some cases, it can be useful to work with other inference products, for example, machine-learning (and grid) representations of marginal likelihoods \citep{stacking,stacking2,Wysocki2020,rift}.
    Additional work is required to define a standardised format for such inference products.
\end{enumerate}

\section*{Acknowledgements}
This work is supported by the Gravitational Wave Data Centre, which is funded under the Astronomy National Collaborative Research Infrastructure Strategy (NCRIS) Program via Astronomy Australia Ltd. (AAL).
This work is supported by the Australian Research Council (ARC) Centre of Excellence CE170100004. PDL is supported by ARC Discovery Project DP22010161. 

\bibliography{refs}

\appendix

\section{Technical details}\label{appendix}
\gwcloud leverages a variety of modern web technologies to provide seamless access via web browsers or an Application Programming Interface (API), exposed to researchers via terminal command line by a publicly available \python client called \texttt{gwcloud-python} (see \url{https://pypi.org/project/gwcloud-python/}).

\subsection{Application Architecture}
In the backend, \gwcloud takes advantage of \texttt{Django}\footnote{\url{https://www.djangoproject.com/}}:  a mature \python-based Model View Controller (MVC)\footnote{\url{https://developer.mozilla.org/en-US/docs/Glossary/MVC}} web framework widely used in the commercial sector\footnote{\url{https://graphql.org/}}.  
Alongside \texttt{Django}, the chosen technology for API transport is \texttt{GraphQL},\footnote{\url{https://graphql.org/}}\footnote{\url{https://docs.graphene-python.org/projects/django/en/latest/}} which provides an efficient and effective way for clients (such as \texttt{gwcloud-python} or any web browser) to request the data they require and only the data they require.  This is in contrast to other less efficient API transport architectures such as Representational State Transfer (REST), which can lead to a variety of issues (e.g., request cascades \footnote{\url{https://leapgraph.com/what-graphql-solves/}}) when dealing with complex data.

In the frontend, the chosen technology is \texttt{React.js}\footnote{\url{https://reactjs.org/}}, which is an industry standard web framework that efficiently handles Document Object Model (DOM) updates to generate fully interactive and dynamic web applications.
To complement \texttt{React.js}, \texttt{Relay.js}\footnote{\url{https://relay.dev/}} (which uses \texttt{GraphQL} fragments and caching to efficiently maintain a consistent state for web applications) makes representation of the data from the web server more efficient. 

\gwcloud exists as one of several projects core to the Gravitational Wave Data Centre (GWDC; see \url{https://gwdc.org.au}): a software engineering initiative of Astronomy Australia Limited (AAL; see \url{https://astronomyaustralia.org.au}), based at Swinburne University of Technology and operated alongside the Astronomy Data and Computing Services (ADACS) team.
ADACS is tasked with providing software services to the Australian astronomy community and the combined resources of it and the GWDC presently consist of approximately fifteen dedicated software development professionals.  

Several of the core GWDC projects are web applications (others include \textsc{GWLab}\footnote{\url{https://gwlab.org.au/}} and \textsc{GWLandscape}\footnote{\url{https://gwlandscape.org.au/}}) with similar infrastructure requirements (e.g., LVK user authentication, execution of ``jobs'' on compute clusters hosting LIGO data; management, visualisation, and searching of these jobs, etc.).
To easily expand and grow these projects and to reduce maintenance overheads\footnote{\url{https://microservices.io/}}, a microservice architecture was chosen for all GWDC web applications.
When a new feature is to be added or a bug is found in an existing service, it is easy to identify and isolate the code involved, reducing complexity and technical debt.
Such isolation also naturally simplifies testing, promoting enhanced reliability and uptime.

Microservice architectures consist of multiple discrete applications running behind the scenes to perform independent and unrelated tasks.
The \gwcloud application, for example, is just a single backend service and frontend \texttt{Javascript} bundle tasked purely with performing tasks related to the submission and management of Bilby jobs.
Other notable services operating in parallel include an authentication service, a database search service and a Job Controller.
The authentication service facilitates integration with the LIGO IDP\footnote{\url{https://ldvw.ligo.caltech.edu/}} and manages accounts of non-LIGO-affiliated users\footnote{\url{https://gwcloud.org.au/auth/}}.
This service also provides details about the user (e.g., LVK membership status, user details such as name and email address, etc.).
The database search service\footnote{\url{https://github.com/gravitationalwavedc/gwcloud_db_search}} provides efficient and powerful job searching based on provided search parameters.
Finally the Job Controller provides a service that \gwcloud and other projects can use to submit and monitor jobs on High Performance Computing (HPC) facilities as well as to fetch the results and files of those jobs while running or once completed\footnote{\url{https://github.com/gravitationalwavedc/gwcloud_job_server}}.

\gwcloud also loosely makes use of microservices in the frontend. The \texttt{React} host is responsible for loading services as required depending on the current URL of the web browser.
This takes advantage of \texttt{Webpack}\footnote{\url{https://webpack.js.org/}} Federated Modules\footnote{\url{https://webpack.js.org/concepts/module-federation/}}.

A limited amount of shared code is present, including the host \texttt{React} module, which is responsible for orchestrating which project to load depending on the currently visited URL.
Other shared code exists between projects, however it is not shared from one location but rather duplicated from a base project template which contains the aforementioned basic core functionality.

\begin{figure*}
    \centering
    \includegraphics[width=0.5\textwidth]{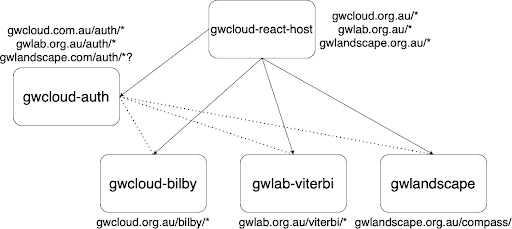}
    \caption{
    Gravitational-wave Data Center Frontend Architecture. 
    This diagram shows how the frontend uses a microservice architecture depending on the URL being visited.
    The \texttt{React} host is always loaded, and is then responsible for loading \texttt{Auth} or application \texttt{javascript} bundles.
    This architecture prevents having a single monolithic \texttt{javascript} bundle that becomes difficult to maintain.
}
    \label{fig:appendix1}
\end{figure*}

\begin{figure*}
    \centering
    \includegraphics[width=0.6\textwidth]{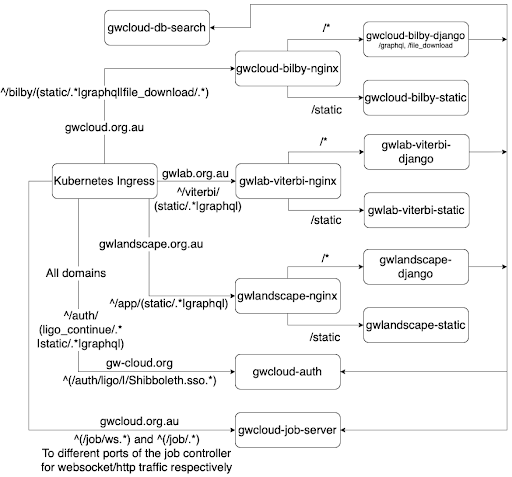}
    \caption{
    Gravitational-wave Data Center Services Architecture; virtual arrangement of backend services.
    All external requests originate from the \texttt{Kubernetes} ingress service and are distributed to the relevant internal services.
    Internal services communicate with other internal services, including the Authentication, Job Server and Database Search services.
}
    \label{fig:appendix2}
\end{figure*}

\subsection{User Experience and Design}
The GWDC implements a Human Centred Design \citep{Norman} approach in the creation of user interfaces (UIs) and client-facing APIs.  In most cases the Design Thinking\footnote{\url{https://www.geeksforgeeks.org/multilevel-queue-mlq-cpu-scheduling/}} [34] variant of Human Centred Design is used.  The main goal of design efforts is to reduce the cognitive load of programming related tasks to allow researchers to focus on scientific challenges.
To our knowledge, this is the first time that Human Centred Design has been intentionally used to improve the UIs and APIs of gravitational wave research applications.

\gwcloud has undergone several Usability Tests \citep{Nielsen,Norman} to inform and validate design choices for the UI and API.  Initial Usability Tests were performed on the UI to develop an understanding of how researchers used the interface and to ultimately build empathy with their needs.
This data was analysed to define the problems researchers faced and to prototype design solutions. Once a solution had been selected it was implemented and then validated with further testing.
This iterative design process is ongoing but has already seen improvement in reducing the number of errors, lowering the barrier of entry, and increasing the user satisfaction, efficiency, and learnability of the UI and API.

\begin{figure*}
    \centering
    \includegraphics[width=0.75\textwidth]{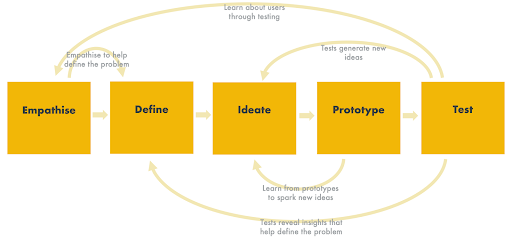}
    \caption{
    Design Thinking Process.  A flowchart showing the steps used by GWDC to develop UIs and APIs.
    This can be a sequential process of empathising with the users, defining the issues they face, ideating solutions, creating a prototype and then testing to determine if the prototype is successful.
    Often insights and data discovered can better inform previous steps leading to an iterative workflow.
}
    \label{fig:appendix3}
\end{figure*}

\subsection{Underlying infrastructure}
To reduce maintenance complexity, \texttt{Kubernetes} is used as the underlying infrastructure for \gwcloud applications.  
\texttt{Kubernetes} is an open source platform designed for containerised applications.
It enables automated operations such as deployments, backups, rollbacks, horizontal virtual resource scaling, name-spaced role-based access control and configuration decoupling from applications\footnote{\url{https://kubernetes.io/docs/concepts/overview/what-is-kubernetes/}}.
Since virtual hosts are abstracted from deployments, applications can be redeployed as needed in an automated self-healing manner\footnote{\url{https://kubernetes.io/docs/concepts/architecture/cloud-controller/}}.

Within \texttt{Kubernetes}, supporting tools are deployed in compliance with the cloud native roadmap\footnote{\url{https://github.com/cncf/trailmap/blob/master/CNCF_TrailMap_latest.pdf}}.
As per the roadmap, all applications involved with \gwcloud are packaged and deployed in the form of Docker containers.
Docker is an Open Container Initiative (OCI)-complaint\footnote{\url{https://docs.docker.com/buildx/working-with-buildx/}} containerisation platform enabling the ingestion of container configurations by other OCI-compliant tools such as \textsc{Podman} or \textsc{Buildah}.
The container images are stored in a container registry.
These container images are then repackaged with default deployment configurations in the form of Helm charts and are stored to a Helm chart repository. 
Both container images and Helm charts are stored within \textsc{Sonatype Nexus}\footnote{\url{https://help.sonatype.com/repomanager3/nexus-repository-administration/formats/docker-registry}}.
As a prerequisite, all sensitive data are declared and initialised within the centralised secrets manager Hashicorp Vault, through its key-value pair secrets engine\footnote{\url{https://www.vaultproject.io/docs/secrets/kv}}.
Values required by deployments from Hashicorp Vault must meet the access requirements configured within Vault\footnote{\url{https://www.vaultproject.io/docs/platform/k8s/injector}}.
The Helm charts are then ingested and deployed to the target \texttt{Kubernetes} cluster through \textsc{ArgoCD}\footnote{\url{https://argo-cd.readthedocs.io/en/stable/operator-manual/architecture/}}: the management tool for the deployment lifecycle of GWDC applications.

\begin{figure*}
    \centering
    \includegraphics[width=0.7\textwidth]{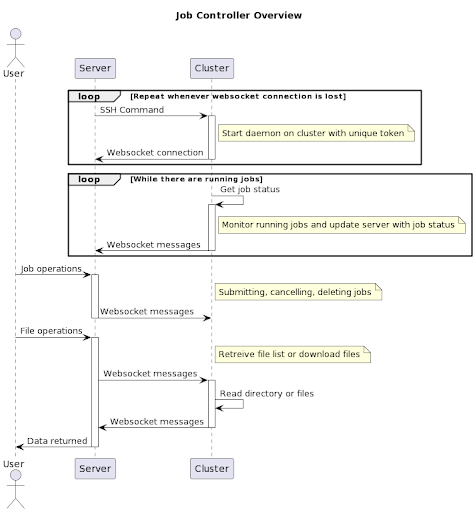}
    \caption{
    High Level Infrastructure Architecture Diagram: \texttt{Kubernetes} cluster component diagram.  
    The component diagram represents a high level segregation of components hosted by the \texttt{Kubernetes} cluster. 
    The virtual hosts represent the virtual machines that are included as part of the \texttt{Kubernetes} cluster. 
    The control plane represents atomic components allowing \texttt{Kubernetes} to be operational.
    Management tools are workloads deployed for managing operations related activities. 
    This involves fundamental requirements such as security, networking, and storage related administration. 
    Application workloads include all custom applications developed for GWDC. At this time of writing, it includes applications directly involved with \gwcloud and \textsc{GWLab}, as another example.
}
    \label{fig:appendix5}
\end{figure*}

\subsection{Interfacing with HPC Facilities}
An underlying component of the GWDC infrastructure underpinning the success of several projects including \gwcloud is the Job Controller: a module responsible for communicating with remote HPC resources such as those based at Swinburne University\footnote{\url{https://supercomputing.swin.edu.au/ }} and Caltech\footnote{\url{https://computing.docs.ligo.org/lscdatagridweb/resources/index.html}}.  
Previously the only similar software solutions have generally been tightly coupled to single clusters; often requiring the cluster’s filesystem to be mounted in a manner allowing web applications to access files directly.
These solutions scale poorly and are hostile to many contemporary practices for HPC facility management.  A new solution was required.

The Job Controller is itself three discrete components: The Job Controller Server\footnote{\url{https://github.com/gravitationalwavedc/gwcloud_job_server}}, Job Controller Client [30], and Bundles\footnote{\url{https://github.com/gravitationalwavedc/gwcloud_job_client}}.  
The Job Controller Server is deployed in the \texttt{Kubernetes} infrastructure and exposes an API that can be used by various modules, including \gwcloud, for submitting jobs; retrieving the status of jobs; cancelling jobs; and retrieving file lists and downloading job files from remote clusters.  
The server is written in \textsc{C++} for multi-threaded performance and uses a \textsc{MySQL} database for persisting information about jobs, their states, and for caching job file lists for complete jobs. 

The Job Controller Client is written in \python and runs as a daemon on all leveraged remote clusters but can be deployed on any system supporting SSH communication with \python 3 installed.
It communicates directly with the server via a \texttt{WebSocket}\footnote{\url{https://developer.mozilla.org/en-US/docs/Web/API/WebSockets_API}} established when the client is initiated, which is instigated by the server via SSH.
The client then forks itself to become a daemon and the SSH connection is dropped.
This architecture has the advantage that the only communication needed by the remote cluster is a brief initiating inbound SSH interaction and \textsc{HTTPS} for any subsequent communication with the server. 
The server can direct the client to submit a new job, cancel a running job, or delete data relating to past jobs.  
The client tracks the state of running jobs via Bundles (described below) and reports job-state updates to the server.
Importantly, the client also provides the ability for the server to request a file list for a job in realtime, and for the server to ask the client to send a job result file over the \texttt{WebSocket} for transfer to a user via browser or API.  A single Job Controller server may have many clients on many remote clusters.

Communication between the client and the server happens over one single \texttt{WebSocket} connection and is scheduled using a Multi Level Priority Queue\footnote{\url{https://www.geeksforgeeks.org/multilevel-queue-mlq-cpu-scheduling/}}: an algorithm taken from operating system design.
This allows higher priority data (such as job file lists) to be sent first over the \texttt{WebSocket}, while lower priority data (such as file transfers) is sent last or as ``best effort.''
This design keeps the client/server communication responsive for real time events (such as when the user requests a file list for a job) at only a slight throughput cost to file transfers, for example.
All communication between the client and server.

If a \texttt{WebSocket} connection is dropped or broken, the client is terminated, and the server attempts to restart the remote client via SSH.
If the SSH connection fails, the server will intermittently retry the SSH connection until it succeeds.  This provides minimal downtime, and resiliency against cluster maintenance or transient connectivity issues between the client and server.

Bundles are the final component of the Job Controller.  They contain the business logic required to prepare a job for submission, submit it, and to check its execution status.  A single Job Controller client may have many bundles, which could represent different projects (\gwcloud, \textsc{GWLab}, etc.), or different versions of runtime codes (e.g. \textsc{Bilby}) leveraged by them.  A versioned history of bundles is maintained by the client to support robust reproducibility of past jobs, if required.  In the case of \gwcloud, its bundle is responsible for rewriting the \texttt{ini} file for the local cluster hosting a job, downloading and storing supporting files, and tracking the state of the job for \textsc{Slurm} and \textsc{Condor} batch schedulers.

\end{document}